\documentclass[%
 reprint,
showpacs,preprintnumbers,
 amsmath,amssymb,
aps,
]{revtex4-1}

\usepackage{graphicx}% Include figure files
\usepackage{dcolumn}% Align table columns on decimal point
\usepackage{bm}% bold math

\usepackage{xcolor}
\usepackage{soul}
\usepackage{amsmath}
\usepackage{verbatim}
\begin{document}

\preprint{APS/123-QED}

\title{Ionic-heterogeneity-induced spiral- and scroll-wave turbulence in mathematical models of cardiac tissue}
\author{Soling Zimik$^1$}
\author{Rupamanjari Majumder$^{2}$}
 \email{rupamanjari.majumder@ds.mpg.de}
\author{Rahul Pandit$^{1}$}
 \email{rahul@iisc.ac.in; also at Jawaharlal Nehru Centre for Advanced Scientific Research, Bangalore, Karnataka, India}
%\altaffiliation[Also at ]{Physics Department, XYZ University.}%Lines break 
\affiliation{%
 $^1$Centre for Condensed Matter Theory, Department of Physics, 
Indian Institute of Science, Bangalore, 560012, India.\\
$^2$Laboratory for Fluid Physics, Pattern Formation and Biocomplexity, Max Planck Institute for Dynamics and Self-Organization,
37077 G\"ottingen, Germany
\\
}%

\date{\today}% It is always \today, today,

\begin{abstract}
Spatial variations in the electrical properties of cardiac tissue can occur because of cardiac diseases. We introduce such gradients into mathematical models for cardiac tissue and then study, by extensive numerical simulations, their effects on reentrant electrical waves and their stability in both two and three dimensions. We explain the mechanism of spiral- and scroll-wave instability, which entails anisotropic thinning in the wavelength of the waves because of anisotropic variation in its electrical properties.
\end{abstract}

\maketitle

%\tableofcontents

%\section{\label{sec:level1}First-level heading}

%%%%%%%%%%%%%%%%%%%%% INTRODUCTION %%%%%%%%%%%%%%%%%%%%%%%%%%%%%
Nonlinear waves in the form of spirals occur in many excitable media, examples of which include Belousov-Zhabotinsky-type systems \cite{zaikin1970concentration}, calcium-ion waves in \textit{Xenopus} oocytes \cite{clapham1995calcium}, the aggregation of \textit{Dictyostelium discoideum} by cyclic-AMP signaling \cite{tyson1989cyclic}, the oxidation of carbon monoxide on a platinum surface \cite{imbihl1995oscillatory}, and, most important of all, cardiac tissue \cite{davidenko1992stationary}. Understanding the development of such spiral waves and their spatiotemporal evolution is an important challenge in the study of extended dynamical systems, in general, and especially in cardiac tissue, where these waves are associated with abnormal rhythm disorders, which are also called arrhythmias. Cardiac tissue can support many patterns of nonlinear waves of electrical activation, like traveling waves, target waves, and spiral and scroll waves~\cite{tyson1988singular}. The occurrence of spiral- and scroll-wave turbulence of electrical activation in cardiac tissue has been implicated in the precipitation of life-threatening cardiac arrhythmias like ventricular tachycardia (VT) and ventricular fibrillation (VF), which destroy the regular rhythm of a mammalian heart and render it incapable of pumping blood. These arrhythmias are the leading cause of death in the industrialized world ~\cite{bayly1998spatial,witkowski1998spatiotemporal,walcott2002endocardial,efimov1999evidence,de1988reentry}. 

Biologically, VF can arise because of many complex mechanisms. Some of these are associated with the development of instability-induced spiral- or scroll-wave turbulence~\cite{fenton2002multiple}. One such instability-inducing factor is ionic heterogeneity~\cite{moe1964computer,jalife2000ventricular}, which arises from variations in the electrophysiological properties of cardiac cells (myocytes), like the morphology and duration of their action-potentials ($AP$s)~\cite{campbell2012spatial,szentadrassy2005apico,
stoll2007spatial,liu1995characteristics}. Such variations may appear in cardiac tissue because of electrical remodeling~\cite{elshrif2015representing,nattel2007arrhythmogenic,
cutler2011cardiac}, induced by alterations in ion-channel expression and activity, which arise, in turn, from diseases~\cite{amin2010cardiac} like ischemia~\cite{harken1978two,jie2010mechanisms}, some forms of cardiomyopathy~\cite{Sivagangabalan14}, and the long-QT syndrome~\cite{viswanathan2000cellular}.  To a certain extent, some heterogeneity is normal in healthy hearts; and it has an underlying physiological purpose~\cite{szentadrassy2005apico,antzelevitch1991heterogeneity,
furukawa1990differences,fedida1991regional,zicha2004transmural,samie2001rectification}; but, if the degree of heterogeneity is more than is physiologically normal, it can be arrhythmogenic~\cite{janse2004electrophysiological,
jie2010mechanisms,nattel2007arrhythmogenic}. It is important, therefore, to explore ionic-heterogeneity-induced spiral- or scroll-wave turbulence in mathematical models of cardiac tissue, which allow us to control this heterogeneity precisely, in order to be able to identify the nonlinear-wave instability that leads to such turbulence. 
We initiate such a study by examining the effects of this type of heterogeneity in three cardiac-tissue models, 
which are, in order of increasing complexity and biological realism, 
(a) the two-variable Aliev-Panfilov model~\cite{aliev1996simple}, (b) the ionically realistic O'Hara-Rudy (ORd) model~\cite{o2011simulation} in two dimensions (2D), and (c) the ORd model in an anatomically realistic simulation domain. In each one of these models, we control parameters (see below) in such a way that the ion-channel properties change anisotropically in our simulation domains, thereby inducing an anisotropic spatial variation in the local action potential duration $APD$. We show that this variation in the $APD$ leads, in all these models, to an anisotropic reduction of the  wavelength of the spiral or scroll waves; and
this anisotropic reduction of the wavelength paves the way for an instability that precipitates turbulence, the mathematical analog of VF, in these models.

%%%%%%%%%%%%%%%%% MATERIALS AND METHODS %%%%%%%%%%%%%%%%%%%%%%%%%%%
%\section{Materials and Methods}
\begin{table}[!ht]
%\caption{{\bf Table of currents.}}
	\begin{tabular}{l l }
	\multicolumn{2}{l}{} \\
	\hline
	\hline
	$\rm{I_{Na}}$ & fast inward $\rm{Na^+}$ current \\
	$\rm{I_{to}}$ & transient outward $\rm{K^+}$ current \\
	$\rm{I_{CaL}}$ & L-type $\rm{Ca^{2+}}$ current \\
	$\rm{I_{Kr}}$ & rapid delayed rectifier $\rm{K^+}$ current \\
	$\rm{I_{Ks}}$ & slow delayed rectifier $\rm{K^+}$ current \\
	$\rm{I_{K1}}$ & inward rectifier $\rm{K^+}$ current \\
	$\rm{I_{NaCa}}$ & $\rm{Na^+/Ca^{2+}}$ exchange current \\
	$\rm{I_{NaK}}$ & $\rm{Na^+/K^+}$ ATPase current\\
	$\rm{I_{Nab}}$ & $\rm{Na^+}$ background current \\
	$\rm{I_{Cab}}$ & $\rm{Ca^{2+}}$ background current\\
	$\rm{I_{pCa}}$ & sarcolemmal $\rm{Ca^{2+}}$ pump current\\
	$\rm{I_{Kb}}$ & $\rm{K^+}$ background current\\
	$\rm{I_{CaNa}}$ & $\rm{Na^+}$ current through the L-type $\rm{Ca^{2+}}$ channel \\
	$\rm{I_{CaK}}$ & $\rm{K^+}$ current through the L-type $\rm{Ca^{2+}}$ channel\\
	\hline
	\end{tabular}
    \caption{The various ionic currents incorporated in the ORd model are tabulated above. The symbols used for the currents follow Ref. \cite{o2011simulation}.}
\label{table}
\end{table}

%%%%%%%%%%%%%%%%%%%%%%%%%%%%%%%%%%%%%%%%%%%%%%%%%%%%%%%%%%%%%%%%%%%%%%%%%%%%
The Aliev-Panfilov model provides a simplified description of an excitable cardiac cell~\cite{aliev1996simple}. It comprises a set of coupled ordinary differential equations (ODEs), for the normalized representations of  the transmembrane potential $V$ and the generalized conductance $r$ of the slow, repolarizing current: 
\begin{eqnarray}
\frac{dV}{dt} & = & -kV(V-a)(V-1)-Vr; %+ I_{ex} ;
\label{ali1}
\end{eqnarray}
\begin{eqnarray}
\frac{dr}{dt} & = & [ \epsilon + \frac{\mu _1 r}{\mu _2 +V}] [ -r-kV(V-b-1)] ;
\label{ali2}
\end{eqnarray} 
fast processes are governed by the first term in Eq.(\ref{ali1}), whereas, the slow, recovery phase of the $AP$ is determined by the function $ \epsilon +
\frac{\mu _1 r}{\mu _2 +V}$ in Eq.(\ref{ali2}). The parameter $a$ represents the threshold of activation and $k$ controls the magnitude of the transmembrane current. We use the standard values for all parameters~\cite{aliev1996simple}, except for the parameter $k$. We write $k$=$g \times k_o$, where $g$ is a multiplication factor and $k_o$ is the control value of $k$. In 2D simulations we introduce a spatial gradient (a linear variation) in the value of $k$ along the vertical direction of the domain.
To mimic the electrophysiology of a human ventricular cell, we perform similar studies using a slightly modified version of the ionically-realistic O'Hara-Rudy model (ORd)~\cite{o2011simulation, zimik2016instability}. Here, the transmembrane potential $V$ is governed by the ODE
\begin{equation}
\frac{d{V}}{d{t}} = -\frac{I_{\rm{ion}}}{C_m}, \hspace{1cm} I_{ion}=\Sigma_x{I_x} ,
\label{ode2.1}
\end{equation}
where $I_x$, the membrane ionic current, 
for a generic ion channel $x$, of a cardiac cell, is
\begin{equation}
I_x = G_x f_1(p_{act}) f_2(p_{inact}) (V_m - E_x),
\end{equation}
where $C_m$=1 $\mu$F is the membrane capacitance, $f_1(p_{act})$ and $f_2(p_{inact})$ are, respectively, functions of probabilities of activation ($p_{act}$) and inactivation ($p_{inact}$) of the ion channel $x$, and $E_x$ is its Nernst potential. We give a list of all the ionic currents in the ORd model in Table~\ref{table}. We write $G_x=g\times{G_{xo}}$, where $G_{xo}$ is the original value of the maximal conductance of the ion channel $x$ in the ORd model, and $g$ is a multiplication factor. We model gradients in $G_x$ as follows: 
\begin{equation}
G_x(y) =  [g_{min}+\frac{y(g_{max}-g_{min})}{L}]G_{xo}, 0\leq y \leq L;
\label{model}
\end{equation}
here, $L$ is the length of the side of the square simulation domain, and $g_{max}$ and $g_{min}$ are, respectively, the maximal and minimal values of $g$;
we can impose gradients in $k$ in the Aliev-Panfilov model in the same manner. For simplicity, we induce the gradient along one spatial direction only: the vertical axis in 2D; and the apico-basal (apex-to-base) direction in 3D. 
The spatiotemporal evolution of $V$ in both % the Aliev-Panfilov and ORd 
models is governed by the following reaction-diffusion equation:
\begin{equation}
\frac{\partial{V}}{\partial{t}}+I=\nabla.(\mathcal{D}\nabla V),
\label{monodomain}
\end{equation} 
where $\mathcal{D}$ is the diffusion tensor, and $I = \frac{I_{\rm{ion}}}{C_m}$ and $kV(V-a)(V-1)+Vr$ for ORd and Aliev-Panfilov models, respectively. For the numerical implementation of the diffusion term in Eq.~(\ref{monodomain}), we follow Refs.~\cite{majumder2016scroll,zimik2016instability}. 
We construct our anatomically realistic simulation domain with processed human-ventricular data, obtained by using Diffusion Tensor Magnetic Resonance Imaging (DTMRI)~\cite{DTMRI}. 
%%%%%%%%%%%%%%%%%%%%%%%%%%%%%%%%%%%%%%%%%%%%%%%%%%%%%%%%%%%%%%%%%%%%%%%%%%%%
%%%%%%%%%%%%%%%%%%% RESULTS & DISCUSSION %%%%%%%%%%%%%%%%%%%

\begin{figure}[!ht]
\includegraphics[width=\linewidth]{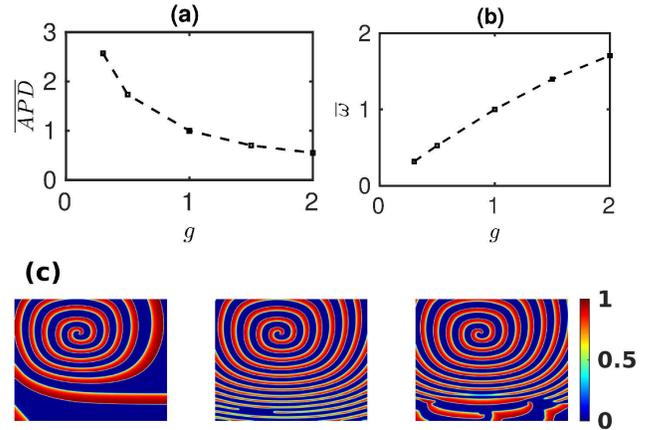}
\caption{(Color online) Variation of (a) $\overline{APD}$ and (b) $\overline{\omega}$ (see text) with $k$. $\overline{APD}=APD/APD_o$; here, $APD_o$ is the control value of $APD$ at $g=1$ (so $\overline{APD}=1$ at $g=1$); we also use other combinations of ($\overline{APD}$, $g$) in our numerical simulations. We find that $\overline{APD}$ decreases with increasing $k$; however, $\overline{\omega}$ increases with increasing $k$. (c) Pseudocolor plots of $V$, at three representative times (time increases from left to right), illustrating the precipitation of the spiral-wave instability in the Aliev-Panfilov model with a linear gradient in $k$; the video S1 in the Supplemental Material~\cite{SI} shows the complete spatiotemporal evolution of this instability.} 
\label{AP}
\end{figure}

In Fig.~\ref{AP}(a) we show the variation, with the parameter $g$, of $\overline{APD}=APD/APD_o$, where $APD_o$ is the control $APD$ value for $g=1$. We find that $\overline{APD}$ decreases with increasing $g$. Changes in the $APD$ at the single-cell level influence electrical-wave dynamics at the tissue level. In particular, such changes affect the rotation frequency $\omega$ of reentrant activity (spiral waves). If $\theta$ and $\lambda$ denote, respectively, the conduction velocity and wavelength of a plane electrical wave in tissue, then
%\begin{equation}
$\omega \simeq \frac{\theta}{\lambda}$,
%\hspace*{0.1\linewidth}
$\lambda \simeq \theta\times APD$.
%\label{CV}
%\end{equation}
Therefore, if we neglect curvature effects~\cite{qu2000origins}, the spiral-wave frequency 
\begin{equation}
\omega\simeq\frac{1}{APD}.
\label{APD_omega}
\end{equation}
We find, in agreement with this simple, analytical estimate, that $\omega$ decreases as the $APD$ increases. We show this in Fig.~\ref{AP}(b) by plotting $\overline{\omega}=\omega/\omega_0$ versus $g$; 
here, $\omega_0$ is the frequency for $g=1$ \footnote{For the parameter $a$ this simple relation between $\omega$ and $APD$ is not observed, because change in $a$ affects not only the APD but also other quantities like $\theta$, which has effects on the value of $\omega$.}.

Similarly, in the ionically realistic ORd model, changes in the ion-channel conductances $G_x$ alter the $APD$ of the cell and, therefore, the spiral-wave frequency $\omega$. In Figs.~\ref{APD_omega} (a1) and (a2) we present a family of plots to illustrate the variation in $\overline{APD}$ with changes in $G_x$. We find that $\overline{APD}$ decreases with an increase in $g$ for most currents ($I_{Kr}$, $I_{Ks}$, $I_{K1}$, $I_{Na}$ and $I_{NaK}$); but it increases for some other currents ($I_{Ca}$, $I_{NaCa}$ and $I_{to}$). The rate of change of $\overline{APD}$ is most significant when we change $G_{Kr}$; by contrast, it is most insensitive to changes in $G_{Na}$ and $G_{to}$. In Figs.~\ref{APD_omega} (b1) and (b2) we show the variation of $\overline{\omega}$ with $g$ for different ion channels $x$. We find that changes in $G_x$, which increase $APD$, decrease $\omega$ and vice versa; this follows from Eq.~(\ref{APD_omega}). The sensitivity of $\omega$, with respect to changes in $G_x$, is most for $G_x=G_{Kr}$ and least for $G_x=G_{to}$: $\overline{\omega}$ \textit{increases} by $\Delta \overline{\omega} \simeq 1.23$, as $g$ goes  from 0.2 to 5; for $G_{to}$, the same variation in $g$ \textit{decreases} the value of $\overline{\omega}$ by  $\Delta \overline{\omega} \simeq 0.04$.
             
We now investigate the effects, on spiral-wave dynamics, of spatial gradients in $k$, in the 2D Aliev-Panfilov model, and in $G_x$, in the 2D ORd model. A linear gradient in $k$, in the Aliev-Panfilov model, induces a gradient in $\overline{\omega}$ (see Fig.~\ref{AP}(b)); and such a spatial gradient in $\overline{\omega}$ induces a spiral-wave instability in the low-$\overline{\omega}$ region. In Fig.~\ref{AP}(c) we demonstrate how a gradient in $k$ ($g_{max}=1.5$ and $g_{min}=0.5$) leads to the precipitation of this instability (also see video S1 of the Supplemental Material~\cite{SI}). 

\begin{figure}[!ht]
\includegraphics[width=\linewidth]{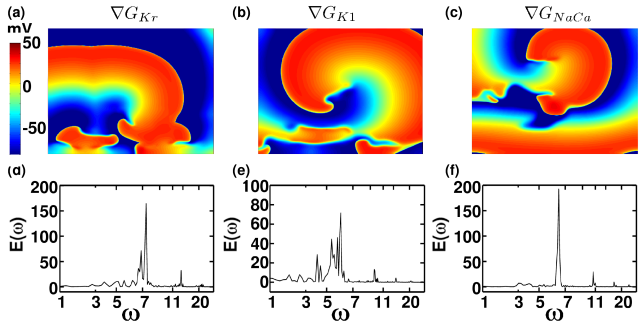}
\caption{(Color Online) Pseudocolor plots of the transmembrane potential $V_m$ illustrating spiral-wave instabilities from our numerical simulations of the 2D ORd model for human ventricular tissue, with spatial gradients in (a) $G_{Kr}$, (b) $G_{K1}$, and (c) $G_{NaCa}$ (because $G_{NaCa}$ decreases with $g$ (Fig.~\ref{APD_omega}), the gradient in $G_{NaCa}$ must be chosen to be the negative of that in Eq.~(\ref{model})); in (a)-(c) the local value of $\omega$ decreases from the top of the simulation domain to its bottom. Power spectra of the time series of $V_m$, from representative points in our simulation domain, are shown for gradients in (d) $G_{Kr}$, (e) $G_{K1}$, and (f)  $G_{NaCa}$; the spectra in (d) and (e) are consistent with the onset of spiral-wave turbulence; the power spectrum in (f) shows the continuation of periodic electrical activity, in spite of wave breaks.}\label{spiral_break} 
\end{figure}

Similarly, for each current listed in Table~\ref{table} for the ORd model, we find wave breaks in a medium with a gradient in $G_x$. %with a spread in the local value of $\omega$ $\Delta \overline{\omega} > 0.12$.
We illustrate, in Fig.~\ref{spiral_break}, such wave breaks in our 2D simulation domain, with a gradient ($\nabla{G_x}$) in any $G_x$, for $3$ representative currents; we select $I_{Kr}$, because it has the maximal impact on the single-cell $APD$, and also on $\omega$ in tissue simulations; and we choose $I_{K1}$ and $I_{NaCa}$, because they have moderate and contrary effects on $APD$ and $\omega$ (Figs.~\ref{APD_omega}). Our results indicate that gradient-induced wave breaks are generic, insofar as they occur in both the simple two-variable (Aliev-Panfilov) and the ionically realistic (ORd) models of cardiac tissue. In Figs.~\ref{spiral_break} (d-f), we present power spectra of the time series of $V$, recorded from a representative point of the simulation domain; these spectra show broad-band backgrounds, which are signatures of chaos, for the gradients $\nabla{G_{Kr}}$ and $\nabla{G_{K1}}$; however, the gradient $\nabla G_{NaCa}$ induces wave breaks while preserving the periodicity of the resultant, reentrant electrical activity, at least at the points from which we have recorded $V$.

\begin{figure}[!ht]
\includegraphics[width=\linewidth]{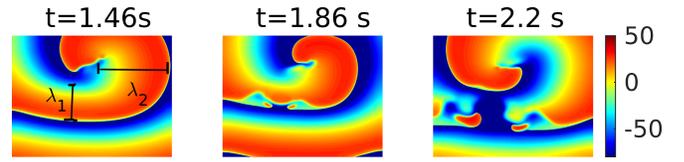}
\caption{(Color Online) Pseudocolor plots of $V_m$ illustrating the development of a spiral-wave instability, with the passage of time $t$, in the 2D ORd model, with a spatial gradient in $G_{NaCa}$. The left frame shows the thinning of the spiral arm ($\lambda$ varies along the spiral arm, and $\lambda_2>\lambda_1$) indicated  just before the spiral wave breaks (see the middle and the right frames).
} \label{mechanism} 
\end{figure}

The instability in spiral waves occurs because spatial gradients in $k$ (Aliev-Panfilov) or in $G_x$ (ORd) induce spatial variations in both $\overline{APD}$ and $\overline{\omega}$: In our simulation domain, the local value of $\overline{\omega}$ ($\overline{APD}$) decreases (increases) from the top to the bottom. In the presence of a single spiral wave (left panel of Fig.~\ref{mechanism}), the domain is paced, in effect, at the frequency $\omega$ of the spiral, i.e., with a fixed time period $T = 1/\omega =APD+DI$, where $DI$ is the diastolic interval (the time between the repolarization of one $AP$ and the initiation of the next $AP$). Thus, the bottom region, with a long $APD$, has a short $DI$ and vice versa. The restitution of the conduction velocity $\theta$ implies that a small $DI$ leads to a low value of $\theta$ and vice versa~\cite{cherry2004suppression} (see Fig. S1 in the Supplemental Material~\cite{SI}). To compensate for this reduction of $\theta$, the spiral wave must reduce its wavelength $\lambda$, in the bottom, large-$APD$ (small-$DI$) region, so that its rotation frequency $\omega \simeq \frac{\theta}{\lambda}$ remains unchanged, as shown in Fig.~\ref{mechanism} (also see video S2 in the Supplemental Material~\cite{SI}), where the thinning of the spiral arms is indicated by the variation of $\lambda$ along the spiral arm ($\lambda_2>\lambda_1$, in the pseudocolor plot of $V_m$ in the top-left panel $t$= 1.46 s). Clearly, this thinning is anisotropic, because of the uni-directional variation in $k$ or $G_x$; this anisotropy creates functional heterogeneity in wave propagation, which leads in turn to the spiral-wave instability we have discussed above (Fig.~\ref{mechanism}). 

In the ORd model, we find that gradients in $G_{Kr}$ easily induce instabilities of the spiral for small values of $\Delta g \equiv g_{max}-g_{min} \simeq 0.5$; by contrast, in a medium with gradients in $G_{to}$, the spiral remains stable for values of $\Delta g$ as large as 4.8. This implies that the stability of the spiral depends on the magnitude of the gradient in $\omega$ that is induced in the medium. 

\begin{figure}[!ht]
\includegraphics[width=\linewidth]{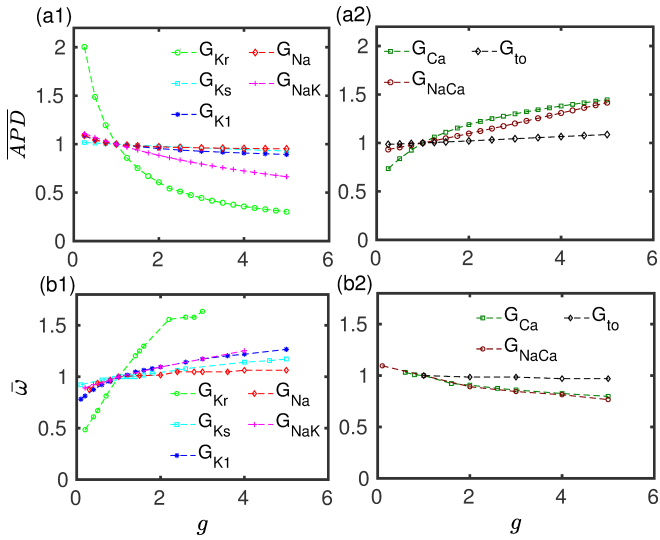}
\caption{(Color online) Plots of $\overline{APD}$ and $\overline{\omega}$ versus $g$; here, $\overline{APD}=APD/APD_o$ and $\overline{\omega}=\omega/\omega_0$, where $APD_o$= 250 ms, and $\omega_0$= 4.38 Hz are, respectively, the control values of $APD$ and $\omega$; (a1) and (a2) show, respectively, that $\overline{APD}$ decreases with the conductances $G_x$, for the currents $I_{Kr}$, $I_{Ks}$, $I_{K1}$, $I_{Na}$ and $I_{NaK}$; however, it increases with increasing $G_x$, for the currents $I_{CaL}$, $I_{NaCa}$ and $I_{to}$; (b1) and (b2) show that the variation of $\overline{\omega}$, with the various channel conductances, is consistent with Eq.~(\ref{APD_omega}).} 
\label{APD_omega}
\end{figure}

% \subsection{Scroll-wave instability}

In Fig.~\ref{scroll} (also see video S3 in the Supplemental Material~\cite{SI}), we extend our study to illustrate the onset of scroll-wave instabilities in a 3D, anatomically realistic human-ventricular domain, in the presence of spatial gradients in $G_{Kr}$. In mammalian hearts, the $APD$ is typically lower in the apical region as compared to that in the basal region~\cite{szentadrassy2005apico}. Therefore, we use values of the $APD$ that increase from the apex to the base (and, hence, $\omega$ decreases from the apex to base). With $g_{max} (G_{Kr})$= 6 and $\Delta g$= 4, we observe breakup in a scroll wave that is otherwise stable in the absence of this spatial gradient. We note that the mechanism for the onset of such scroll-wave instabilities is the same as in 2D, and it relies on the gradient-induced anisotropic thinning of the scroll wavelength.
 \begin{figure}[!ht]
 \includegraphics[width=\linewidth]{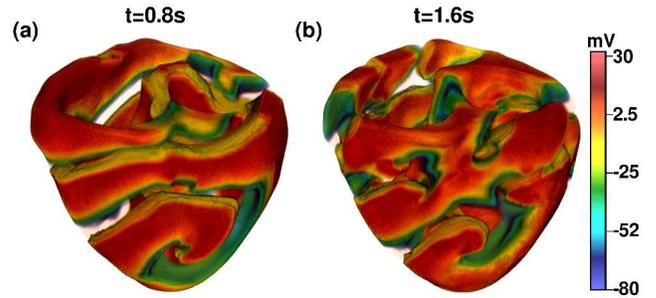}
 \caption{(Color Online) Scroll-wave instabilities in our anatomically realistic human-ventricular domain, in the presence of an apico-basal gradient in $G_{Kr}$} \label{scroll} 
 \end{figure}

%\section{Discussion}

We have shown that gradients in parameters that affect the $APD$ of the constituent cells induce spatial gradients in the local value of $\omega$. This gradient in the value of $\omega$ leads to an anisotropic reduction in the wavelength of the waves, because of the conduction-velocity restitution property of the tissue, and it paves the way for spiral- and scroll-wave instability in the domain. This gradient-induced instability is a generic phenomenon because we obtain this instability in the simple Aliev-Panfilov and the detailed ORd model for cardiac tissue. Such an instability should be observable in any excitable medium that has the conduction-velocity-restitution property. We find that the spiral or scroll waves always break up in the low-$\omega$ region. This finding is in line with that of the experimental study by Campbell, \textit{et al.},~\cite{campbell2012spatial} who observe spiral-wave break-up in regions with a large $APD$ in neonatal-rat-ventricular-myocyte cell culture. We find that the stability of the spiral is determined by the magnitude of the gradient in $\omega$; the larger the magnitude of the gradient in the local value of $\omega$, the more likely is the break up of the spiral or scroll wave. By using the ORd model, we find that $\omega$ varies most when we change $G_{Kr}$ (as compared to other ion-channel conductances) and, therefore, spiral waves are most unstable in the presence of a gradient of $G_{Kr}$. By contrast, we find that $\omega$ varies most gradually with $G_{to}$, and hence the spiral wave is most stable in the
presence of a gradient in $G_{to}$ (as compared to gradients in other conductances). 

Earlier studies have investigated the effects of ionic-heterogeneity on
spiral-wave dynamics. The existence of regional ionic heterogeneities have been
found to initiate spiral waves~\cite{defauw2013initiation}, attract spiral
waves to the heterogeneity~\cite{defauw2014small}, and destabilize spiral
waves~\cite{xu1998two}. The presence of $APD$ gradients in cardiac tissue has
been shown to drive spirals towards large-$APD$ (low $\omega$) regions~\cite{ten2003reentry}. A study by Zimik, {\it et al.},~\cite{zimik2016instability} finds that spatial gradients in $\omega$, induced by gradients in the density of fibroblasts, can precipitate a spiral-wave
instability. However, none of these studies provides a clear understanding of the mechanisms underlying the onset of spiral- and scroll-wave instabilities, from a fundamental standpoint. Moreover, none of these studies has carried out a detailed calculation of the pristine effects of each individual major ionic currents, present in a myocyte, on the spiral-wave
frequency; nor have they investigated, in a controlled manner, how gradients in
ion-channel conductances lead to spiral- or scroll-wave instabilities. Our work makes up for these lacunae and leads to specific predictions that should be tested experimentally. 

\begin{acknowledgments}
We thank the Department of Science and Technology (DST), India, and the Council for Scientific and Industrial Research (CSIR), India, for financial support, and Supercomputer Education and Research Centre (SERC, IISc) for computational resources.
\end{acknowledgments}

\bibliography{references}

\end{document}

\maketitle

\section*{Video Captions:}
Video S1: {\bf Spiral-wave instability in the Aliev-Panfilov model.} Video of pseudocolor plots of transmembrane potential $V$ showing the formation of spiral-wave instability in a medium with gradient in $k$: $g_{min}$=0.5 and $g_{max}$= 1.5. For the video, we use 10 frames per second with each frame separated from the succeeding frame by 20ms in real time.\\

\noindent Video S2: {\bf Spiral-wave instability in the ORd model.} Video pseudocolor plots of transmembrane potential $V_m$ showing the formation of spiral-wave instability in a medium with a gradient in $G_{Naca}$ ($g_{min}$=0.2 and $g_{max}$= 2). For the video, we use 10 frames per second with each frame separated from the succeeding frame by 20ms in real time.\\

\noindent Video S3: {\bf Scroll-wave instability.} Video pseudocolor plots of transmembrane potential $V_m$ showing the formation of scroll-wave instability in an anatomically-realistic model for human ventricles. A linear gradient in $G_{Kr}$ is applied along the apico-basal direction: $g_{min}=2$ in the apex and $g_{max}$=6 in the base. For the video, we use 10 frames per
second with each frame separated from the succeeding frame by 20ms in real time.\\

\section*{Figures:}
\begin{figure}[!ht]
\includegraphics[width=\linewidth]{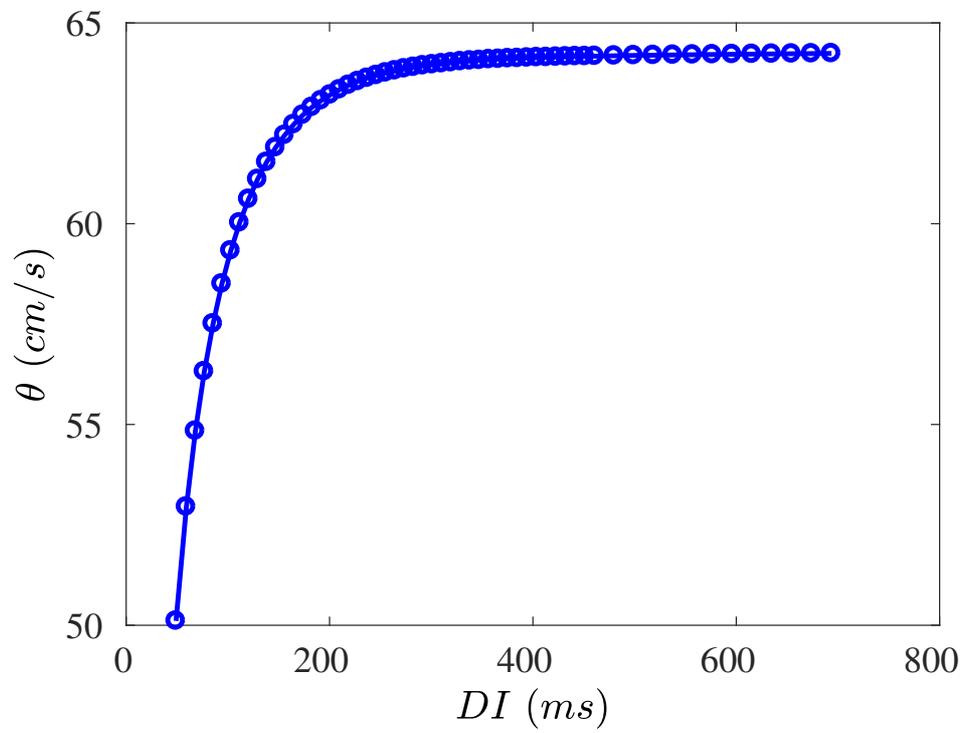}
\caption{Figure showing a conduction-velocity restitution curve, generated by using the ORd model. The value of conduction velocity $\theta$ initially increases with the increase of diastolic interval $DI$ and saturates at large values of $DI$.}
\end{figure}